\documentclass[a4paper,12pt]{article}
\usepackage{a4wide}
\usepackage{graphicx}
\usepackage{epstopdf}
\usepackage{epsfig}
\usepackage[sumlimits,intlimits,namelimits]{amsmath}
\usepackage{amssymb,color}
\usepackage[english]{babel}

\numberwithin{equation}{section}
\usepackage{cite}

\begin{document}
\begin{titlepage}
\begin{flushright}
TUM-HEP-723/09\\
SI-HEP-2009-09 \\[0.2cm] 
\today
\end{flushright}

\vspace{1.2cm}
\begin{center}
\Large\bf\boldmath
Sequential Flavour Symmetry Breaking
\unboldmath
 \end{center}

\vspace{0.5cm}
\begin{center}
{\sc Thorsten Feldmann,$\,^a$ Martin Jung$\,^{b,c}$ 
 and Thomas Mannel$\,^{b}$}

\vspace{0.7cm}

{\small 
${}^a$ {\sf Physik Department T31, Technische Universit\"at M\"unchen,\\ 
 D-85747 Garching, Germany.}
\vspace{1\baselineskip}

${}^b$ {\sf Theoretische Physik 1, Fachbereich Physik,
Universit\"at Siegen,\\ D-57068 Siegen, Germany.}
\vspace{1\baselineskip}

${}^c$ {\sf
Instituto de F\'{\i}sica Corpuscular,
CSIC-Universitat de Val\`encia, 
Apartado de Correos 22085,\\
E-46071 Valencia, Spain.}
\vspace{1\baselineskip}
}

\end{center}

\vspace{3em}
\begin{abstract}

The gauge sector of the Standard Model (SM) exhibits 
a flavour symmetry which allows for independent unitary
transformations of the fermion multiplets.
In the SM the flavour symmetry is broken by the Yukawa couplings
to the Higgs boson, and the resulting fermion masses and mixing
angles show a pronounced hierarchy. In this work we connect
the observed hierarchy to a sequence of intermediate effective
theories, where the flavour symmetries are broken in a step-wise
fashion by vacuum expectation values of suitably constructed
spurion fields. We identify the possible scenarios
in the quark sector and discuss some implications of this approach.

\end{abstract}

\end{titlepage}

\newpage
\pagenumbering{arabic}

%papercontent
%\newpage
%\tableofcontents
%\newpage

\section{Introduction}
The origin of flavour remains one of the main mysteries in modern particle physics,
and many attempts have been made to understand the phenomenon of flavour by
postulating certain (discrete) flavour symmetries
 (see e.g.\ 
\cite{Ma:2006sk,Antusch:2007re,Blum:2007jz,Varzielas:2008kr,King:2008zz,Bazzocchi:2008sp}
 and references therein), or 
by localizing fermions in extra dimensions (see e.g.\ 
\cite{ArkaniHamed:1999dc,Gherghetta:2000qt,Grossman:1999ra,Huber:2000ie,Agashe:2004ay,Davidson:2007si,Csaki:2008eh,Buras:2009ka} 
and references therein), 
to name two popular ideas. While such scenarios can 
successfully explain some of the issues related to the hierarchies observed in 
fermion masses and mixings, the origin of the proposed new mechanisms (e.g.\ 
from the embedding into a Grand Unified Theory or even String Theory, respectively) 
still remains an open issue.

Alternatively, we may start from a bottom-up approach in which the phenomenon of 
flavour is just parametrized as in the Standard Model (SM). In fact, the SM
has an approximate global flavour symmetry $G_F$ (see below), which is broken by the Yukawa 
couplings, inducing the fermion masses and mixings. 
Such an explicit symmetry breaking is usually 
parametrized by introducing spurion fields with a definite behaviour under the symmetry 
to be broken. In the case at hand, focusing on the quark sector,
the Yukawa matrices $Y_U$ and  $Y_D$ are
considered as complex spurion fields \cite{D'Ambrosio:2002ex}, transforming non-trivially
under $G_F$. 

A special role is played by the top quark, which has a Yukawa coupling of
order one, breaking the original flavour symmetry group $G_F$ to a smaller
sub-group $G_F'$ (see below), which is still a good symmetry
as long as the remaining Yukawa couplings are negligible. 
In a recent paper \cite{Feldmann:2008ja}, two of us have shown 
that in such a case it is convenient to consider a non-linear representation of $G_F$ in 
which the subgroup $G_F^\prime$ is  linearly realized. 
In this context, it turned out to be useful to assign a canonical mass dimension to 
the Yukawa spurion fields, since in this way the 
top Yukawa coupling could be understood as originating from 
a dimension-four operator while the remaining Yukawa  terms 
are dimension five, thereby reflecting  the hierarchy between 
the top mass and the lighter quark masses.\footnote{A similar construction
can be performed in the lepton sector, when the SM is minimally extended
by a dimension-5 operator in order to describe non-vanishing neutrino masses
\cite{Cirigliano:2005ck,Feldmann:2008av}.} 
If we take this approach seriously, two immediate implications arise:
\begin{itemize}
 \item The spontaneous breaking $G_F \to G_F'$ induces 
       Goldstone modes who call for a dynamical interpretation.
       One possibility is to consider \emph{local} flavour symmetries,
       where Goldstone modes become the longitudinal modes for
       massive gauge bosons.
       Another alternative is to keep the Goldstone modes as physical
       axion-like degrees of freedom.\footnote{An option to avoid Goldstone
       modes alltogether is to restrict oneselves to \emph{discrete}
       flavour symmetries.} 
       These issues will be discussed in somewhat more detail
       in a separate publication \cite{Feldmann:2009albrecht}.

 \item The breaking $G_F \to G_F'$ induced by the top-quark Yukawa coupling
       can be considered the first step in a sequence of flavour symmetry
       breaking steps taking place at different physical scales $\Lambda \gg \Lambda'
       \gg \Lambda'' \gg \ldots$ \, Through the VEVs of the spurion fields, the
        hierarchy of scales should be directly related to the observed hierarchy
        for quark masses and mixings. 
\end{itemize}
In the following, we shall identify the flavour sub-groups for each of the
intermediate effective theories in the construction above and identify
the corresponding representations for quark fields, spurions and 
Goldstone modes. We will also briefly discuss the requirements
for the spurion potential necessary for such a scenario.

\section{Successive flavour symmetry breaking}

In this section we identify the sequence of intermediate (residual)
flavour symmetries which arise when the original flavour symmetry
of the SM gauge sector is broken in a step-wise fashion
at different scales, set by the VEVs of the relevant
spurion fields and linked to the observed hierarchies in the
quark masses and CKM angles. 
Considering the Yukawa sector for the quarks,
\begin{align}
 - {\cal L}_Y & = Y_U \, \bar Q_L \, \tilde H \, U_R + Y_D \, \bar Q_L \, H \, D_R 
    + \mbox{h.c.}  \,,
\end{align}
we may consider independent phase transformations for the 3 quark multiplets 
($Q_L,U_R,D_R$) and the Higgs field ($H$).
Among these 4 phases, 2 are identified as baryon number $U(1)_B$ and weak hypercharge $U(1)_Y$
which are not broken by the Yukawa matrices, whereas the 2 remaining $U(1)$ symmetries are
broken by $\langle Y_U \rangle \neq 0 $ or $\langle Y_D \rangle \neq 0$.
We thus define the flavour group in the quark sector as\footnote{Our discussion
differs from the one in \cite{D'Ambrosio:2002ex} where the independent phase rotations for
the Higgs fields have been overlooked.} 
\begin{align}
 G_F &= SU(3)^3 \times U(1)^4/(U(1)_B\times U(1)_Y)  \cr 
   & = SU(3)_{Q_L} \times SU(3)_{U_R} \times SU(3)_{D_R} \times U(1)_{U_R} \times U(1)_{D_R} \,.
\label{GFquark}
\end{align}
For the Yukawa sector to be formally invariant under $G_F$, we 
assign the following transformation properties to the spurion fields, 
\begin{equation}
 Y_U \sim (3,\bar 3,1)_{-1,0}  \,, \qquad Y_D \sim (3,1,\bar 3)_{0,-1} \,,
\end{equation}
where the terms in brackets refer to the three $SU(3)$ factors, and the subscripts
to the two $U(1)$ factors, respectively.
Counting parameters, we have $2 \times 18=36$ entries for the spurions $Y_{U,D}$ and
$3\times 8 + 2 = 26$ symmetry generators, leaving $36-26 = 10$ physical parameters
in the quark Yukawa sector, which can be identified with the 6 quark masses, the
3 CKM angles, and the CP-violating CKM phase (see also \cite{Berger:2008zq}).\footnote{%
%%%%%%%%%%%%%%%%%%%%%%%%%%%%%
Similarly, considering the $U(1)$ phases 
in the lepton sector, we obtain the SM flavour group 
$$
  G_F^{\rm lepton} = U(3)^2/(U(1)_e\times U(1)_\mu \times U(1)_\tau)
$$
for massless neutrinos, and
$$
  \tilde G_F{}^{\rm lepton} = U(3)^2
$$
for massive neutrinos which are generated by a lepton-number violating
dim-5 term in the Lagrangian
$$
 -{\cal L}_{\rm Maj} = \frac{1}{\Lambda_{\rm L}} \, g_\nu \, (H \ell_L)^T (H \ell_L) \,.
$$
In the first case, we have 18 parameters in the spurion $Y_E$ and $2 \times 8 -1 =15$
symmetry generators, leaving 3 physical parameters to be identified with the 3 charged
lepton masses.
In the second case, we have $18 + 12=30$ parameters from the spurions $Y_E$ and $g_\nu$,
from which we subtract $2\times 9 = 18$ symmetry generators, to obtain 12 physical parameters,
which are the 6 lepton masses, the 3 PMNS angles, one Dirac phase, and the 2 Majorana phases.
%%%%%%%%%%%%%%%%%%%%%%%%%%%%%
}

\begin{table}[t!pb]
 \caption{\label{tab:mq} \small
 SM values for the quark masses \cite{Amsler:2008zz}, and approximate scaling with
 the Wolfenstein parameter $\lambda \sim 0.2$. Light-quark masses (u,d,s)
  are given in the $\overline{\rm MS}$ scheme at $\mu=2$~GeV, 
  charm and bottom masses as $\bar m_c(\bar m_c)$ and $\bar m_b(\bar m_b)$,
  and the top mass is evolved down to the scale $m_b$. The evolution between
   the scales $m_b$ and $m_c$ is negligible for our considerations.}
\begin{center}
\begin{tabular}{l | c c c}
\hline
  &  u & d & s  \\
\hline
$m_q$ & $[1.5 - 4.5]$~MeV & $[5.0 - 8.5]$~MeV & $[80-155]$~MeV \\
$n_q=\log_{\lambda}(m_q/m_t)$ &
        $6-9$ & $6-8$ & $4-6$  \\
\hline \hline
& c & b & t \\
\hline
$m_q$ &$[1.0 - 1.4]$~GeV & $[4.0 - 4.5]$~GeV & $[250-300]$~GeV \\
$n_q=\log_{\lambda}(m_q/m_t)$ &$3-4$ & $2-3$ & 0 \\
\hline
\end{tabular}
\end{center}
\end{table}

In order to specify the sequence of flavour symmetry breaking,
we have to identify a hierarchy between the Yukawa
entries $(Y_U)_{ij}$ and $(Y_D)_{ij}$. However, before the flavour 
symmetry is actually broken, the Yukawa matrices can be freely rotated 
by transformation matrices in $G_F$, and therefore the 
a-priori ranking of individual entries in the Yukawa matrices seems to be
somewhat ambiguous. On the other hand, the right-handed rotations 
and a \emph{common} left-handed rotation for up- and down-quarks are
not observable in the SM, anyway, leaving the quark masses and CKM
angles as the only relevant parameters.
We therefore find it sufficient to choose a basis where
the right-handed rotations are unity, while for the left-rotations
we restrict ourselves to matrices $V_{u_L}$ and $V_{d_L}$ which
scale in the same manner as the CKM matrix.
This leaves us with the generic power counting\footnote{During the sequence of
flavour symmetry breaking, some of the entries can actually be
set to zero by exploiting the freedom to rotate the VEVs of
certain spurion fields with respect to the corresponding residual 
flavour group.}
\begin{align}
 \langle Y_U\rangle_{ij} 
 & \sim  (V_{u_L})_{ij} \, (y_u)_j 
  \sim \left( \begin{array}{ccc} 
\lambda^{n_u} & \lambda^{1+n_c} & \lambda^{3}  \\
\lambda^{1+n_u} & \lambda^{n_c} & \lambda^{2} \\
\lambda^{3+n_u} & \lambda^{2+n_c} & 1 
\end{array} \right)
\,, \cr
 \langle Y_D\rangle_{ij} 
& \sim  (V_{d_L})_{ij} \, (y_d)_j 
\sim \left( \begin{array}{ccc} 
\lambda^{n_d} & \lambda^{1+n_s} & \lambda^{3+n_b}  \\
\lambda^{1+n_d} & \lambda^{n_s} & \lambda^{2+n_b} \\
\lambda^{3+n_d} & \lambda^{2+n_s} & \lambda^{n_b} 
\end{array} \right) \,,
\end{align}
where we introduced the scaling for quark
Yukawa couplings with the Wolfenstein parameter 
($\lambda \sim 0.2 \ll 1 $) as $y_i \sim \lambda^{n_i}$ 
(with $n_t=0$),
and inserted the standard power counting for CKM elements,
\begin{align}
V_{u_L} \sim V_{d_L} \sim V_{\rm CKM} \sim \left( 
\begin{array}{ccc} 
1 & \lambda&\lambda^3 \\
\lambda & 1 & \lambda^2 \\
\lambda^3 & \lambda^2 & 1
\end{array} \right) \,.
\end{align}
The scaling of the quark masses can be
constrained from the phenomenological information
in Table~\ref{tab:mq}, where we assume in the following that 
renormalization-group effects (in the sequence of effective theories
to be constructed) do not change the hierarchies observed at low scales 
a lot.
More precisely, to keep the discussion simple,
we restrict ourselves to:
\begin{itemize}
 \item $n_d > n_s > n_b>0$ and $n_u > n_c> n_t \equiv 0$,
 \item $n_c \geq n_b$ and $n_s > n_c$.
\end{itemize}
The remaining degree of freedom in choosing values for the $n_i$
leads to several options,
among which are also cases where one or two spurions receive
their VEV at the same scale simultaneously. 
To be concrete, we focus on three cases with more or
less natural and distinct scale separation,
(a1) $n_c < n_b+2 < n_b+3 < n_s $,
(a2) $n_c < n_b+2 < n_s < n_b+3$, and
(b)  $n_b +2 < n_c < n_s < n_c+1$, 
which are summarized in Table~\ref{tab:summ}.
A detailed derivation of the various steps in the 
flavour symmetry breaking can be found in the appendix.

\label{sec:succ}

\begin{table}[t!!hb]
 \caption{\label{tab:summ} \small
Three alternative sequences of flavour symmetry breaking, and associated 
parameter counting for the Yukawa matrices. Notice that the following 
equalities always hold: 
$$
 \mbox{\# Spurions}+\mbox{\# VEVs}-\mbox{\# Symmetries}=10\,,
$$
$$
 \mbox{\# Goldstones}+\mbox{\# Spurions}+\mbox{\# VEVs}=36\,,
$$
where 10 refers to the 6 quark masses + three CKM rotations + one CKM phase,
and 36 refers to the original $2\times 18$ real parameters in the Yukawa matrices
$Y_U$ and $Y_D$.}
\begin{center} \small
 \begin{tabular}{|c |c | l | c c c c|c|}
\hline \multicolumn{3}{|c|}{Flavour Symmetry} & GBs & Spur. & VEVs & Symm. & Scale \\
\hline\hline
\multicolumn{3}{|l|}{$SU(3)_{Q_L}\times SU(3)_{U_R} \times SU(3)_{D_R} \times U(1)^2$}
& 0 & 36 & 0 & 26 & \\
 \multicolumn{3}{|l|}{  $SU(2)_{Q_L}\times SU(2)_{U_R} \times SU(3)_{D_R} \times U(1)^3$ }
& 9 & 26 & 1 & 17& $\Lambda \sim y_t \, \Lambda$  \\
  \multicolumn{3}{|l|}{$SU(2)_{Q_L}\times SU(2)_{U_R} \times SU(2)_{D_R} \times U(1)^3$}
& 14 & 20 & 2 & 12& $\Lambda' \sim y_b \, \Lambda$ \\
\hline \hline
(a)   &  \multicolumn{2}{|l|}{$ SU(2)_{D_R} \times U(1)^4$}
& 19 & 14 & 3 & 7& $\Lambda^{(2a)} \sim y_c \, \Lambda$ \\
\hline
& (a1)  & $ SU(2)_{D_R} \times U(1)^3$ & 20 & 12 & 4 & 6 
& $\Lambda^{(3a1)} \sim y_b \lambda^2 \, \Lambda$ 
\\
&   & $ SU(2)_{D_R} \times U(1)^2$ & 21 & 11 & 5 & 5 
& $\Lambda^{(4a1)} \sim y_b \lambda^3 \, \Lambda$ 
\\
&   & $U(1)^2$ & 24 & 6 & 6 & 2 
& $\Lambda^{(5a1)} \sim y_s \, \Lambda$ 
\\
\hline 
& (a2)  & $ SU(2)_{D_R} \times U(1)^3$ & 20 & 12 & 4 & 6 
& $\Lambda^{(3a1)} \sim y_b \, \lambda^2 \, \Lambda$ 
\\
&   & $ U(1)^3$ & 23 & 8 & 5 & 3 
& $\Lambda^{(4a2)} \sim y_s \, \Lambda$ 
\\
&   & $U(1)^2$ & 24 & 6 & 6 & 2 
& $\Lambda^{(5a2)} \sim y_b \lambda^3 \, \Lambda$ 
\\
\hline \hline
 (b)  &  \multicolumn{2}{|l|}{$SU(2)_{U_R} \times SU(2)_{D_R} \times U(1)^3$}
& 17 & 16 & 3 & 9 & $\Lambda^{(2b)} \sim y_b \lambda^2 \, \Lambda$ 
\\
      &  \multicolumn{2}{|l|}{$SU(2)_{D_R} \times U(1)^3$}
& 20 & 12 & 4 & 6 & $\Lambda^{(3b)} \sim y_c \, \Lambda$ 
\\
      &  \multicolumn{2}{|l|}{$U(1)^3$}
& 23 & 8 & 5 & 3 & $\Lambda^{(4b)} \sim y_s \, \Lambda$ 
\\
      &  \multicolumn{2}{|l|}{$U(1)^2$}
& 24 & 6 & 6 & 2 & $\Lambda^{(5b)} \sim y_c\lambda \, \Lambda$ 
\\
\hline \hline
\multicolumn{3}{|l|}{  $  U(1)^2$  \hfill (CP\hspace{-1em}{\large \bf / }) }
& 24 & 4 & 7+1 & 2 & $\Lambda^{(6)} \sim y_s \lambda \, \Lambda$
\\
\multicolumn{3}{|l|}{  --  \hfill (CP\hspace{-1em}{\large \bf / }) }
& 26 & 0 & 9+1 & 0 & $\Lambda^{(7)} \sim y_{u,d} \, \Lambda$ \\
\hline
 \end{tabular}

\end{center}

\end{table}

Let us discuss some common and distinct features of the different scenarios:
\begin{itemize}
\item Common to all scenarios is the second step of symmetry breaking
  which (at least in our set-up with only one electroweak Higgs doublet)
  is unambiguously induced by the VEV for the $(Y_D)_{33}$ element which
  gives rise to the bottom-quark mass.
  Below the scale $\Lambda' \sim y_b \Lambda$, the residual flavour symmetry
  is
 \begin{align}
  G_F'' = U(2)_{Q_L} \times U(2)_{U_R} \times U(2)_{D_R} \,.
\label{GFpp}
 \end{align}
  At first glance, it appears as just the 2-family analogue of
  the original flavour group $G_F$. However, there are two important
  differences: First, it appears one additional $U(1)$ factor compared
  to $G_F$. Second, it still contains an off-diagonal spurion field
  $\chi_s$, which is a doublet of $SU(2)_{Q_L}$ and the only spurion
  which is charged under the additional $U(1)$. Only if this spurion field
  (and the associated breaking of the extra $U(1)$ symmetry) 
  were absent, we would recover
  an effective 2-family model where, as is well-known, one would have
  no CP-violation in the quark Yukawa sector. 

\item From an aesthetic point of view, the alternative labeled (a1)
in Table~\ref{tab:summ} is somewhat
favoured. It can be realized with a rather natural hierarchy of scales. For instance
taking
$$  n_b=2 \,, \quad n_c=3 \,, \quad n_s = 6 \,, \quad n_{u,d} =8\,, $$
which fits well to the phenomenological mass spectrum,
one obtains an equal separation of scales,\footnote{For comparison,
scenario (a2) can be realized, for instance, by $n_b=2.5$, $n_c=3.5$, $n_s=5$,
$n_{u,d}=7$,
leading to the tower of scales $(\lambda^{2.5},\lambda^{3.5},\lambda^{4.5},\lambda^5,
 \lambda^{5.5},\lambda^{6},\lambda^7)\, \Lambda$. Similar, case (b) could be
realized by $n_b=2$, $n_c=4.5$, $n_s=5$, $n_{u,d}=7$ with
$(\lambda^2,\lambda^4,\lambda^{4.5},\lambda^5,\lambda^{5.5},\lambda^6,\lambda^7)\,\Lambda$.} 
$$
 \Lambda^{(n)} = \lambda^{(n+1)} \, \Lambda \,.
$$

Moreover, the smallest non-abelian sub-group
for this case is given by 
$$
  SU(2)_{D_R} \times U(1)^2 \,.
$$
This residual flavour symmetry may thus be taken as the simplest 
non-trivial example to study the dynamics of flavour spurions
and its consequences for flavour physics, including the
construction of higher-dimensional operators for flavour transitions
with minimal flavour violation (or beyond \cite{Feldmann:2006jk}), 
the dynamics of Goldstone modes, and
the construction of realistic scalar potentials.

\item In all cases, the symmetry is eventually broken down to
$$
 U(1)^2 = U(1)_{u_R} \times U(1)_{d_R} \,.
$$
The corresponding effective theory now still contains three complex spurion
fields, among which one spurion is uncharged under either of the two
$U(1)$ groups. Consequently, when the latter acquires its VEV, its phase
cannot be rotated away by symmetry transformation.\footnote{Alternatively,
in a previous step of the construction, one could have identified
two spurion fields with the same quantum numbers, whose VEVs
in general cannot be made real simultaneously. This mechanism
thus gives a particular realization of spontaneous CP-violation \cite{Lee:1973iz}.}
At this very step, we
therefore generically encounter a CP-violating phase, 
which in our case is associated with the $(Y_D)_{12}$ element.

\item Finally, the two $U(1)$ symmetries will be broken by the 
 $(Y_U)_{11}$ and $(Y_D)_{11}$ elements associated with the up- and 
down-quark mass. Notice that these symmetries are chiral, and the
corresponding $U(1)$ anomalies contribute to the effective $\theta$-parameter
in QCD. The related spurion fields may 
serve as a solution to the strong CP problem as in the general Peccei-Quinn
setup \cite{Peccei:1977ur,Weinberg:1977ma,Wilczek:1977pj}.
This will be discussed in more detail in \cite{Feldmann:2009albrecht}.

\end{itemize}

\section{Invariants and Potentials for scalar spurion fields}
In this section we consider how the sequential symmetry breaking, described in
the last section, could be achieved spontaneously. The question of how an appropriate
potential could look like is discussed in many different contexts
(see e.g.\ \cite{Slansky:1981yr,Li:1973mq,Ruegg:1980gf}), but no general recipe
for constructing a potential that leads to a specific symmetry breaking 
has been found. 

In any case, a potential for the spurion fields can only depend on invariants 
under the flavour symmetry group $G_F$. Due to the form of the potential these 
invariants should take the appropriate VEVs which finally 
specify the ten physical parameters (6 quark masses and 4 CKM parameters). 
Of course we are unable to derive a potential which achieves this complicated 
symmetry breaking, but we may at least identify ten independent invariants in terms of 
which we may express the physical quantities. These invariants can be constructed from 
monomials of the basic scalar spurion fields $Y_U(x)$ and $Y_D(x)$, and may thus 
be classified by their canonical dimension.

Before considering the 3-family case, it is instructive to look at the
simpler example of two families with the flavour symmetry
$g_F = SU(2)_{Q_L} \times SU(2)_{U_R} \times SU(2)_{D_R} \times U(1)^2$, first.
It exhibits 11 symmetry generators which leaves 5 physical parameters
(4 masses and the Cabibbo angle) from the 16 parameters in the Yukawa matrices.
Classifying the invariants by increasing canonical dimension, we find
\begin{align}
  i_1^{(2)} &= {\rm tr}(U)\,,
   && v_1^{(2)}/\Lambda^2  = y_u^2 + y_c^2  \,,
 \cr 
  i_2^{(2)} &= {\rm tr}(D)\,,    
   && v_2^{(2)}/\Lambda^2  = y_d^2 + y_s^2   \,,
 \cr  
  i_1^{(4)}  &= {\rm tr}(U^2) - (i_1^{(2)})^2 \,, 
   && v_1^{(4)}/\Lambda^4  = -2 y_u^2 y_c^2  \,,
 \cr 
  i_2^{(4)}  &= {\rm tr}(U D) - i_1^{(2)} i_2^{(2)} \,,  
  && v_2^{(4)}/\Lambda^4   = \sin^2 \theta  (y_c^2 - y_u^2) (y_d^2 - y_s^2) - 
 y_c^2 y_d^2 - y_u^2 y_s^2 \,,
\cr 
  i_3^{(4)}  & = {\rm tr}(D^2) - (i_2^{(2)})^2  \,, 
  && v_3^{(4)}/\Lambda^4  = -2 y_d^2 y_s^2 \, , 
\end{align} 
where we introduced the combinations
\begin{equation}
U=Y_U Y_U^\dagger\,,  \qquad D = Y_D Y_D^\dagger \,,
\end{equation} 
which transform homogeneously under $SU(2)_{Q_L}$,
and where we denote with $v_\alpha^{(k)} =  \langle i_\alpha^{(k)} \rangle$ the VEVs of 
the 5 invariants.
The potential $V = V(i_\alpha ^{(m)})$
may now be expanded around its minimal value in the form
\begin{equation}
V = \sum_{k,m} \sum_{\alpha,\beta} 
\, \frac{1}{\Lambda^{m+k-4}}  
\left(i_\alpha^{(m)} - v_\alpha^{(m)} \right) M^{(m,k)}_{\alpha,\beta} 
   \left(i_\beta^{(k)} - v_\beta^{(k)} \right) \,,
\label{VGF2}
\end{equation}
where $\Lambda$ is a UV-scale which renders the positive semi-definite
matrix $M^{(m,k)}_{\alpha,\beta}$ dimensionless. Notice that higher-dimensional
operators appear unavoidably if we assign canonical mass dimension to the
(scalar) spurion fields $Y_{U,D}$. As already mentioned, the mechanism how
such an effective potential could be generated by integrating out some new degrees
of freedom in an underlying theory, remains an open issue.

In principle we may also invert the relations to obtain the Cabibbo angle and the masses as
functions of the $  v_i^{(k)} $, however, the above invariants are not yet very suitable
for the further discussion:
\begin{itemize}
 \item As we have seen in the previous section, the order of
  the different symmetry-breaking steps depend on the relative size of the Yukawa
  entries, which in the 2-family case are characterized by the exponents $\{n_u,n_c,n_d,n_s,(1+n_s)\}$ (in the hierarchical limit). It is therefore
  desirable to consider invariants that feature the very same exponents.

 \item To put the invariants on a similar footing, they should have the same
        canonical dimension (i.e.\ we have to introduce rational functions of
        the above invariants).

 \item Instead of $i_2^{(4)}$ it would be desirable to have an invariant that
       vanishes in the no-mixing case ($\theta=0$). Such invariants can be
       constructed from the commutator $[U,D]$,
\begin{align}
 i_1^{(8)} &= \det\left( [U,D] \right)\,, 
 && v_1^{(8)}/\Lambda^8 = \frac14 \, (y_c^2-y_u^2)^2 \, (y_s^2-y_d^2)^2 \, \sin^22\theta \,.
\end{align}
\end{itemize}
We therefore modify the above definitions as follows,
\begin{align}
  I_1 &= {\rm tr}(U)\,,
   && V_1/\Lambda^2  = y_u^2 + y_c^2  \,,
 \cr
  I_2 &= {\rm tr}(D)\,,    
   && V_2/\Lambda^2  = y_d^2 + y_s^2   \,,
 \cr   
  I_3 &= \frac12 \left(I_1 - {\rm tr}(U^2)/I_1 \right) \,, 
   && V_3/\Lambda^2  = \frac{y_u^2 y_c^2}{y_u^2 + y_c^2}  \,,
 \cr 
  I_4 & = \frac12 \left( I_2 - {\rm tr}(D^2)/I_2 \right)  \,, 
  && V_4/\Lambda^2  = \frac{ y_d^2 y_s^2}{y_s^2+y_d^2} \, ,
 \cr 
  I_5 &= 4 \, \frac{\det \left([U,D]\right)}{I_1 \, I_2 \, (I_1+I_2)}
 \,,  && V_5/\Lambda^2  = 
 \frac{(y_c^2-y_u^2)^2 \, (y_s^2-y_d^2)^2 \, \sin^22\theta}
  {(y_u^2 + y_c^2)(y_d^2 + y_s^2)(y_u^2 + y_c^2+y_d^2 + y_s^2)} \,.
\label{inv2}
\end{align} 
The invariants $I_{1-5}$ now take their VEVs according to the
power-counting for masses and mixing angles. For instance, 
with our standard case, $n_c < n_s < 1+n_s < n_u \sim n_d$,
we have
$$
  V_1 \sim \lambda^{2n_c} \ \gg \
  V_2 \sim \lambda^{2n_s} \ \gg \
  V_5 \sim \lambda^{2+2n_s} \ \gg \
  V_{3,4} \sim \lambda^{2 n_{u,d}} \,,
$$
which defines the sequence of symmetry breaking.
We may then solve (\ref{inv2}) for masses and mixing angle to obtain
\begin{align}
 y_{c,u}^2 & = \frac{ V_1 \pm \sqrt{V_1 (V_1 - 4 V_3)}}{2\Lambda^2}
 \simeq \Big\{ \begin{array}{l}
                V_1/\Lambda^2 \\ V_3/\Lambda^2
               \end{array}
\,,
\cr 
y_{s,d}^2 & = \frac{ V_2 \pm \sqrt{V_2 (V_2 - 4 V_4)}}{2\Lambda^2}
 \simeq \Big\{ \begin{array}{l}
                V_2/\Lambda^2 \\ V_4/\Lambda^2
               \end{array}
\,,
\cr 
\sin^22\theta &= \frac{(V_1+V_2) \, V_5}{(V_1-4 V_3)(V_2 - 4 V_4)} 
 \simeq \frac{V_5}{V_2} \,,
\end{align}
where the approximate relations refer to the SM hierarchies.

We note in passing that models based on texture zeros, 
which imply relations between 
the masses and the mixing angles  \cite{Fritzsch:1979zq},
 may be mapped onto relations between invariants. 
In turn, a relation between invariants always characterizes a class of Yukawa matrices
which may or may not feature texture zeros in a particular flavour basis.
This may be explicitly demonstrated by considering a simple two-family model with one texture 
zero. 
We use the basis in which $Y_U$ is diagonal and
\begin{equation}
Y_D = \left( \begin{array}{cc}  0 & a \\ a & 2b \end{array} \right) 
\end{equation}
is given in terms of two parameters $a$ and $b$. 
This model implies the relation
\begin{equation}
 V_5 = \frac{4 \, (V_1-4 V_3) \, (\sqrt{V_2 V_4}- 2 V_4)}{V_1+V_2}
    \simeq 4 \sqrt{V_2 V_4} \,,
\end{equation}
which translates into a  relation between the Cabibbo
angle and the down-type masses, 
\begin{equation}
\tan \theta \simeq \sqrt{\frac{m_d}{m_s}} 
\end{equation} 
which is phenomenologically reasonable.

We now turn to the 3-family case, which can be studied along the same lines. 
We have to identify in total ten independent invariants. 
The two quadratic and the three quartic invariants are again given by 
\begin{align}
 & i_1^{(2)} = {\rm tr}(U)\,,
 \qquad  
 i_2^{(2)} = {\rm tr}(D) \,.  %\quad \mbox{ with } \quad     v_2^{(2)}  = m_d^2 + m_s^2 + m_b^2   \,,
\end{align}
and
\begin{align}
  i_1^{(4)} & = {\rm tr}(U^2) - (i_1^{(2)})^2 
\,,  
\quad  
  i_2^{(4)} = {\rm tr}(U D) - i_1^{(2)} \, i_2^{(2)} \,,
 \quad 
  i_3^{(4)}  =  {\rm tr}(D^2) - (i_2^{(2)})^2 \,.
\end{align}
The remaining 5 invariants, which are necessary to specify the physical quark flavour parameters,
thus have to be built from even higher-dimensional invariants. 
For the dimension-6 terms, we choose 
\begin{align}
  i_1^{(6)} & = {\rm tr}(U^3)  - \frac{3}{2} \, i_1^{(4)} i_1^{(2)} -    ( i_1^{(2)} )^3  
 \equiv 3 \, \det(U)
\,,  
\cr 
  i_2^{(6)} &=  
{\rm tr}(U^2 D) - \frac12 \, i_1^{(4)}  i_2^{(2)} - i_2^{(4)}  i_1^{(2)} - i_2^{(2)} \, (i_1^{(2)})^2 
\,, 
\cr  
  i_3^{(6)}  &= 
{\rm tr}(U D^2) - \frac12 \, i_3^{(4)}  i_1^{(2)} - i_2^{(4)}  i_2^{(2)} - i_1^{(2)} \, (i_2^{(2)})^2  \,,
\cr 
  i_4^{(6)}  &= {\rm tr}(D^3)  - \frac{3}{2} \, i_3^{(4)}  i_2^{(2)} -  ( i_2^{(2)} )^3 
 \equiv  3\, \det(D) \,.
\end{align}
Finally, among the dimension-8 invariants only one is linearly independent,
and we choose 
\begin{equation}
  i_1^{(8)} =  {\rm tr}\left(U [U,D] D\right)  \,,
\end{equation} 
which completes the list of invariants for the $3\times 3$ case.

The potential $V = V(i_\alpha^{(m)})$ can again be expanded
as in (\ref{VGF2}).
The sequential breaking of $G_F$ as proposed in the last section can emerge only 
through a hierarchy of VEVs for the various invariants.
This hierarchy has to be put in by hand in (\ref{VGF2}) and may perhaps
find its explanation in an underlying theory above the scale $\Lambda$.\footnote{We
note, however, that restricting ourselves to the most general set of
dimension-4 operators, where
$$
 V = \sum_{i}  m_i^2 \, i_i^{(2)} 
     + \sum_{i,j}2 \rho_{ij} \, i_i^{(2)} i_j^{(2)}  
     +  \sum_{i} \lambda_i \, i_i^{(4)} \,,
$$
 only part of the flavour symmetry will be broken by 
the minimum of the potential, including the case $G_F \to G_F'$ for a particular
sub-set of parameter space.}
In fact, our choice of VEVs is such, that the first breaking  
of $G_F \to G_F^\prime$ is obtained, if the potential $V$ generates a 
(sizeable) VEV for the $i_1^{(2)}$ invariant, only, 
\begin{equation}
v_1^{(2)}  \simeq y_t^2  \Lambda^2 \, , \qquad v_k^{(m)}  \simeq 0 \quad \mbox{otherwise} \,, 
\end{equation}
in which case we obtain a non-vanishing top quark Yukawa coupling, while
all other parameters (which give rise to the lighter quark masses and CKM parameters) 
still (approximately) vanish. 

The next step is the breaking of $G_F^\prime \to G_F^{\prime \prime}$. 
Clearly the relevant potential $V^\prime$ can only depend on the 
invariants of $G_F^\prime$, which we denote as $j_k^{(m)}$. 
As before, we introduce quadratic terms which transform under $SU(2)_L \times U(1)_T$,
namely 2 triplets,
\begin{align}
 U' = Y_U^{(2)} Y_U^{(2)\dagger} \,, \quad
 D' = \tilde Y_D^{(2)} \tilde Y_D^{(2)\dagger} \,,
\end{align}
one charged doublet
\begin{align}
 X' = \tilde Y_D^{(2)} \xi_b \,,
\end{align}
and one singlet
\begin{align}
 \Xi'= \xi_b^\dagger \xi_b \,.
\end{align}
In terms of these, the invariants of dimension-2 can be written as 
\begin{align}
j_1^{(2)} & = {\rm tr} ( U' )  \,,
\quad 
j_2^{(2)}  = {\rm tr} ( D') \,,
\quad 
j_3^{(2)}  =  \Xi' \,,
\end{align}
while the fourth-order invariants are 
\begin{align} 
j_1^{(4)}  & =  
{\rm tr} \left( (U')^2 \right) 
  - ( j_1^{(2)} )^2 \,, 
\qquad
j_2^{(4)}   =  {\rm tr} ( U' D') 
  -    j_1^{(2)}  j_2^{(2)}  \,,  
\cr
j_3^{(4)}  & =  {\rm tr} \left( (D')^2\right) -  ( j_2^{(2)} )^2  \,,
\qquad 
 j_4^{(4)}   =    X'{}^\dagger X' - j_2^{(2)} j_3^{(2)} \,.
\end{align}  
Finally, there are 2 linear independent invariants of dimension 6,
\begin{align}
 j_1^{(6)} &= X'{}^\dagger \, U'\, X' \,, \qquad
 j_2^{(6)} = X'{}^\dagger \, D'\, X' \,.
\end{align}

At tree level, the potential $V^\prime$ simply follows from
the original potential $V$ by expressing the invariants $i_k^{(m)}$
by the invariants $j_\alpha^{(m)}$ and the VEV for the top Yukawa
coupling, see appendix~\ref{app:jinv}. Including radiative corrections
in the effective theory below the scale $\Lambda$ (or more precisely,
below the mass scale of the scalar degree of freedom related to the VEV
$y_t \Lambda$), the parameters of the effective potential might change
accordingly. The general form is thus again given by
\begin{equation}
V^\prime = \sum_{k,m} \sum_{\alpha,\beta} 
\, \frac{1}{(\Lambda')^{m+k-4}}  
\left(j_\alpha^{(m)} - w_\alpha^{(m)} \right) N^{(m,k)}_{\alpha,\beta } 
\left(j_\beta^{(k)} - w_\beta^{(k)} \right) \,.
\label{VGFp}
\end{equation}
The next step in the symmetry breaking, $G_F^\prime \to G_F^{\prime\prime}$,
will then be achieved by $w_3^{(2)} \simeq y_{b}^2 \, (\Lambda^\prime)^2$. This
scheme can be repeated until the complete flavour symmetry is broken.

Note, that the invariants $i_i^{(m)}$ and $j_i^{(m)}$ 
introduced above are all real. 
Therefore, the parameters $M^{(m,k)}_{i,j}$ and
$N^{(m,k)}_{i,j}$ have to be real as well to yield a hermitian potential.
As described above, the CKM phase, corresponding to the SM mechanism for CP-violation, 
appears when one of the spurion fields receives a complex VEV.
The potential allows for spontaneous CP-violation, as soon as 
an invariant of one of the residual flavour symmetries becomes
complex. In the scenarios discussed above, this is the case for
\begin{eqnarray}
G_F^{3a1}: && L^{(4)}_1 = 
\mbox{Re}\left(\chi^*_{13} \, \xi_d^\dagger \, \xi_s \, \chi_{23} \right)\,,\qquad 
L^{(4)}_2 = 
\mbox{Im}\left(\chi^*_{13} \, \xi_d^\dagger \, \xi_s \, \chi_{23}\right)\,,
\\
\mbox{and} \quad G_F^{(3b)}: && L'^{(4)}_1 = 
\mbox{Re}\left(\xi_u^\dagger\,\xi_c\,\xi_s^\dagger\,\xi_d\right)\,,
\qquad \quad 
L'^{(4)}_2 = \mbox{Im}\left(\xi_u^\dagger\,\xi_c\,\xi_s^\dagger\,\xi_d\right)\,,
\end{eqnarray}
where $L^{(')(4)}_2$ is odd under CP.

As in the 2-family example, we again introduce rational functions of
the invariants that are convenient for the discussion of power-counting
or parameter relations in models with texture zeros. The modified set
of invariants for the 3-family case reads
\begin{align}
  I_1 &= {\rm tr}(U)\,,
   && V_1/\Lambda^2  = y_u^2 + y_c^2 + y_t^2  \sim \lambda^0 \,,
 \cr
  I_2 &= {\rm tr}(D)\,,    
   && V_2/\Lambda^2  = y_d^2 + y_s^2 + y_b^2 \sim \lambda^{2n_b} \,,
 \cr   
  I_3 &= \frac12 \left(I_1 - {\rm tr}(U^2)/I_1 \right) \,, 
   && V_3/\Lambda^2  = \frac{y_u^2 y_c^2+ y_u^2 y_t^2 + y_c^2 y_t^2}{y_u^2 + y_c^2+y_t^2}
 \sim \lambda^{2n_c}  \,,
 \cr 
  I_4 & = \frac12 \left( I_2 - {\rm tr}(D^2)/I_2 \right)  \,, 
  && V_4/\Lambda^2  = \frac{ y_d^2 y_s^2 + y_d^2 y_b^2 + y_s^2 y_b^2}{y_s^2+y_d^2+y_b^2} 
 \sim \lambda^{2n_s} \,,
 \cr 
  I_5 &= \det(U)/I_1/I_3 \,,
    && V_5/\Lambda^2 = \frac{y_u^2 y_c^2 y_t^2}{y_u^2 y_c^2+ y_u^2 y_t^2 + y_c^2 y_t^2} 
 \sim \lambda^{2n_u} \,,
 \cr 
  I_6 &= \det(D)/I_2/I_4 \,,
    && V_6/\Lambda^2 = \frac{y_d^2 y_s^2 y_b^2}{y_d^2 y_s^2+ y_d^2 y_b^2 + y_s^2 y_b^2}
 \sim \lambda^{2n_d} 
\,,
\label{inv3a}
\end{align}
which determines the 6 Yukawa couplings corresponding to the quark masses, and
\begin{align}
  I_7 &=  \frac{{\rm tr} \left(U[U,D]D\right)}{I_1 \, I_2 \, (I_1+I_2)}
 \,,  && V_7/\Lambda^2  \simeq y_b^2 \,  \theta_{23}^2  \sim \lambda^{2(n_b+2)} \,,
\cr 
  I_{8} &= \frac12 \, \frac{\det\left( \left[U,\left[U,D \right]\right] \right)}
  {I_1^2 I_2 (I_1+I_2)^2 I_3^2 I_7}
 \,, && V_8/\Lambda^2 \simeq   y_b^2  \theta_{13}^2
    + y_s^2  \frac{\theta_{12} \theta_{13}}{\theta_{23}}   \cos\delta 
 \sim  \lambda^{2 (n_b+3)} + \lambda^{2 (n_s+1)}  \,,
\cr 
  I_{9} &= \frac12 \, \frac{\det\left( \left[\left[U,D \right],D\right] \right)}
  {I_2^2 I_1 (I_1+I_2)^2 I_4^2 I_7 }
 \,, && V_9/\Lambda^2 \simeq  y_b^2  \left( \theta_{13}^2 + \theta_{12}^2 \theta_{23}^2 - 2 \theta_{12} \theta_{23} \theta_{13}  \cos\delta \right)
\sim \lambda^{2(n_b+3)}
\,,
\cr
 I_{10} &= -\frac{i}{2} \, \frac{\det\left( \left[U,D \right] \right)}{I_1^2 I_2^2 (I_3+I_4)} \,,
&& V_{10}/\Lambda^2 \simeq y_s^2 \theta_{12} \theta_{23} \theta_{13}  \sin\delta
\sim \lambda^4 \, \lambda^{2 (n_s+1)}
\label{inv3b}
\end{align}
which determines the angles and the CP-violating phase 
in the standard parametrization \cite{PDG}. Again, the invariants $I_{7-10}$
are defined in such a way that they vanish in the no-mixing case. Moreover,
$I_{10} \neq 0$ signals CP-violation.

We may again solve for the SM parameters to obtain
the quark Yukawa couplings $y_{t,c,u}^2(V_{1,3,5})$
and $y_{b,s,d}^2(V_{2,4,6})$,
as well as the (approximate) solutions for the mixing angles
\begin{align}
 \theta_{23}^2  &\simeq \frac{V_7}{V_2} \,, \qquad
 \theta_{13}^2  \simeq \frac{V_8}{V_2}
%+ {\cal O}\left(\frac{y_s^2}{y_b^2 \lambda^4}\right)  
\,, 
\cr 
 \left(\theta_{12} \, \cos\delta - \frac{\theta_{13}}{\theta_{23}}\right)^2  & \simeq 
 \frac{V_9}{V_7} - \frac{V_2^2 V_{10}^2}{ V_4^2 V_7 V_8}
 \,, \qquad 
 \theta_{12}^2 \, \sin^2\delta \simeq \frac{V_2^2 V_{10}^2}{ V_4^2 V_7 V_8} \,,
\end{align}
where we also neglected terms of order $\lambda^{-4} \, y_s^2/y_b^2$.
Finally, we consider again a simple model with texture zeros
in the $3\times 3$ Yukawa matrices \cite{Fritzsch:2002ga},
\begin{align}
 Y_U = \left( \begin{array}{ccc}
       0 & C_u & 0
\\
  C_u^* & 0 & B_u 
\\
  0 & B_u^* & |A_u |       
              \end{array}
\right)
\,, \qquad
 Y_D =\left( \begin{array}{ccc}
       0 & C_d & 0
\\
  C_d^* & 0 & B_d 
\\
  0 & B_d^* & |A_d |       
              \end{array}
\right)
\,,
\end{align}
which yields the following approximate relations between quark masses
and mixing angles
\begin{align}
 \frac{|V_{ub}|^2}{|V_{cb}|^2} & \simeq \frac{\theta_{13}^2}{\theta_{23}^2} \simeq \frac{m_u}{m_c}
\,,
\qquad 
 \frac{|V_{td}|^2}{|V_{ts}|^2}  \simeq \frac{\theta_{13}^2 + \theta_{12}^2 \theta_{23}^2 - 2 \theta_{12} \theta_{23} \theta_{13}  \cos\delta}{\theta_{23}^2} \simeq \frac{m_d}{m_s} \,.
\end{align}
As before, this can be formulated in a basis-independent way in
terms of the following approximate relations between invariants
\begin{align}
  & \frac{V_8}{V_7} \simeq \sqrt\frac{V_5}{V_3} \,,
\qquad
   \frac{V_9}{V_7} \simeq  \sqrt\frac{V_6}{V_4} \,.
\end{align}

\section{Conclusions} 

In this paper we have shown how the hierarchies in quark masses and 
mixings can be associated with a particular sequence of flavour symmetry breaking.
The different scales at which the individual steps of partial
flavour symmetry breaking occur are separated among each other 
by not more than 1-2 orders of magnitude. Depending on the assumed
power counting for the quark masses, we have identified different
scenarios that are compatible with phenomenology. We have also
given some general arguments for the possible form of scalar
potentials that may realize the sequence of flavour symmetry
breaking and identified the invariants that may be used to
expand the potential around its minimum or to classify ans\"atze
for the Yukawa matrices involving texture zeros 
in a basis-independent way.

In all cases, the minimal \emph{non-abelian} flavour sub-group is 
given by $SU(2)_{D_R} \times U(1)^{2(3)}$.
Its further breaking eventually leads to an effective theory with a 
residual $U(1)^2$ flavour symmetry, where one of the spurion fields
is uncharged. When this spurion achieves a complex VEV, its phase 
cannot be rotated away and provides the one and only source for CP-violation 
in the quark Yukawa sector. The CP-violating phase is thus generated
at rather low scales (compared to, say, a GUT scale).

A dynamical interpretation of the Goldstone modes,
appearing at each step of the (global) flavour symmetry breaking, 
can be achieved by promoting the flavour symmetries to local ones,
where the Goldstone modes become the longitudinal modes of the
corresponding massive gauge bosons. One the other hand, 
the final chiral $U(1)^2$ symmetries are anomalous and the associated
Goldstone bosons couple to the QCD instantons. They may thus
be used to resolve the strong CP-problem as in the general Peccei-Quinn
setup, with the corresponding Goldstone modes appearing as axion fields. 
Details will be presented in a separate publication \cite{Feldmann:2009albrecht}.

\subsection*{Note added}
While completing this work, the paper \cite{Kagan:2009bn} 
appeared, where a 2-Higgs-doublet scenario
with a large ratio of VEVs ($\tan\beta \sim m_t/m_b \gg 1$)
was considered. In this case, the original flavour symmetry
is broken in one step as $G_F \to G_F'' = U(2)^3$, see Eq.~(\ref{GFpp})
and the discussion in \cite{Feldmann:2008ja}. 
The possible enhancement with $\tan\beta$ allows for interesting
observable deviations from the SM and from minimally flavour-violating scenarios with
minimal Higgs sector. In \cite{Kagan:2009bn} it has been shown 
that they can be identified in a very transparent
way using the non-linear representation of flavour symmetries suggested
in \cite{Feldmann:2008ja}. It is evident, that the related change in the 
hierarchies of the Yukawa matrices for $\tan\beta \gg 1$ 
would also imply a different pattern for
the sequence of flavour symmetry breaking which could be worked out 
in an analogous way as presented in our work.

\subsection*{Acknowledgements}
We would like to thank Michaela Albrecht for many helpful discussions. 
TM wants to thank Aneesh Manohar for a helpful discussion on the invariants. 
This work has been supported in part by
the German Research Foundation (DFG, Contract 
No.~MA1187/10-1) and by the German Ministry of Research (BMBF,
Contract No.~05HT6PSA),  by the EU
    MRTN-CT-2006-035482 (FLAVIAnet), by MICINN
 (Spain) under grant
     FPA2007-60323, and by the Spanish Consolider-Ingenio 2010
     Programme CPAN (CSD2007-00042).

\begin{appendix}

\section{Sequence of flavour-symmetry breaking in the SM}

In this appendix, we present the detailed derivation of the
different scenarios for sequential flavour-symmetry breaking
as discussed in the text.

\subsection{Leading order}

Neglecting all terms of ${\cal O}(\lambda)$ in $Y_U$ and $Y_D$, only the top-quark Yukawa
coupling in $(Y_U)_{33}$ survives, due to our general assumption $n_q > 0$. 
We thus obtain the breaking (which -- apart from the additional $U(1)$ factors -- coincides with the discussion in \cite{Feldmann:2008ja})
\begin{align}
G_F \to G_F' & = SU(2)_{Q_L} \times SU(2)_{U_R} \times SU(3)_{D_R} \times U(1)_T \times U(1)_{U_R^{(2)}} \times U(1)_{D_R}  
\label{lam:1}
\\
  & \sim  SU(2)_{Q_L} \times SU(2)_{U_R} \times SU(3)_{D_R} 
   \times U(1)_{Q_L^{(2)}} \times U(1)_{U_R^{(2)}} \times U(1)_{D_R} \,,
\label{lam:2}
\end{align}
where the equivalence in the second line arises if we take into account
the globally conserved baryon number, implying the relation
 $$
3 B = T + Q_L^{(2)} + U_R^{(2)} + D_R
$$ for the quark charges,
where $T$ counts the quark-number for the third generation in $Q_L$ and $U_R$, and 
$Q_L^{(2)}$ and $U_R^{(2)}$ for the first two generations
(see also appendix~\ref{app:u1}).
The decomposition of $Y_U$ and $Y_D$ in terms of irreducible representations of $G_F'$
and the representation of the 9 Goldstone modes 
($\Pi_{L,U_R}^{a=4..8}$, $\Pi_{L}^8=-\Pi_{U_R}^8$) remains as in 
\cite{Feldmann:2008ja}, with 
\begin{align}  
 Y_U & = {\cal U}(\Pi_L) \, \left( 
\begin{array}{cc} Y_U^{(2)} &
  \begin{array}{c} 0 \\ 0
  \end{array} \\
  \begin{array}{cc} 0 & 0
  \end{array} & y_t \, \Lambda
\end{array} \right) {\cal U}^\dagger(\Pi_{U_R})
\,,\quad 
Y_D  = {\cal U}(\Pi_L) \, \left( 
\begin{array}{c}
 \tilde Y_D^{(2)} \\
 \xi_b^\dagger
\end{array} \right) 
\,,
\label{decomp1}
\end{align}
and ${\cal U}(\Pi)= \exp\left[i \Pi^a T^a/\Lambda\right]$.

\subsection{Order $\Lambda'/\Lambda$}

Let us first consider the transformation properties
of the residual spurion fields with respect to $G_F'$
(here the subscripts refer to the $U(1)$ factors defined in
Eq.~(\ref{lam:1})), and their scaling with $\lambda$,
\begin{align}
Y_U^{(2)} & \sim (2,2,1)_{1,-1,0} 
\ \propto \ \left( \begin{array}{cc} 
\lambda^{n_u} & \lambda^{1+n_c}  \\
\lambda^{1+n_u} & \lambda^{n_c}  
\end{array} \right) \Lambda \,, 
\cr 
\tilde Y_D^{(2)} & \sim (2,1,\bar 3)_{1,0,-1} 
\ \propto \ 
\left( \begin{array}{ccc} 
\lambda^{n_d} & \lambda^{1+n_s} & \lambda^{3+n_b} \\
\lambda^{1+n_d} & \lambda^{n_s} & \lambda^{2+n_b} 
\end{array} \right) \Lambda \,, 
\cr 
\xi_b^\dagger & \sim (1,1,\bar 3)_{0,0,-1}
\ \propto \ \left( \begin{array}{ccc} 
\lambda^{3+n_d} & \lambda^{2+n_s} & \lambda^{n_b}
\end{array} \right) \Lambda \,.
\end{align}

We now assume that at the scale $\Lambda' \ll \Lambda$,
the next-highest entry in the residual spurion fields gets
its VEV. For $n_c > n_b$, 
the spurion $\xi_b^\dagger$ will have the largest eigenvalue,
\footnote{If we allow for $n_c = n_b$, the spurion $Y_U^{(2)}$ also will get its
VEV simultaneously, such that in the scenario (a) discussed below, the scales
$\Lambda'$ and $\Lambda''$ would coincide.}
\begin{align} 
 \langle \xi_b^\dagger \rangle = (0,0, \tilde y_b) \, \Lambda \equiv 
  (0,0,x_b) \, \Lambda' \,, 
  \label{xidr} 
\end{align}
with $x_b = {\cal O}(1)$ such that $\tilde y_b \sim \Lambda'/\Lambda \sim m_b/m_t$.
Similarly as for the discussion of the 2HDM with large $\tan\beta$ in
\cite{Feldmann:2008ja}, this further breaks the flavour symmetry to
\begin{align}
G_F' \to G_F'' & = 
SU(2)_{Q_L} \times SU(2)_{U_R} \times SU(2)_{D_R} \times U(1)_{Q_L^{(2)}} \times U(1)_{U_R^{(2)}} \times U(1)_{D_R^{(2)}} 
\label{lamp:1}
\\
& \sim 
SU(2)_{Q_L} \times SU(2)_{U_R} \times SU(2)_{D_R} \times U(1)_{\rm III} \times U(1)_{U_R^{(2)}} \times U(1)_{D_R^{(2)}} \,,
\label{lamp:2}
\end{align}
where now $U(1)_{\rm III}$ acts on all quarks in the third generation.
The 5 additional Goldstone modes ($\Pi'_{D_R}{}^{a=4..8}$) 
are introduced as 
\begin{align}
 \tilde Y_D^{(2)} & = \left( 
\begin{array}{cc}
 Y_D^{(2)} & \chi_s
\end{array} \right)  {\cal U}^\dagger(\Pi'_{D_R}) \,,
\cr 
\xi_b^\dagger & = \left( 
\begin{array}{ccc}
 0 & 0 & x_b \Lambda'
\end{array} \right)  {\cal U}^\dagger(\Pi'_{D_R}) \,.
\end{align}
with ${\cal U}(\Pi') = \exp\left[ i \Pi'{}^a T^a/\Lambda' \right]$.

%%%%%%%%%%%%%%%%%%%%%%%%%%%%%%%%%%%%%%%%%%%%%%%%%%%%%%%%%%%%%%%%%%%%%%%%%%%%%%

\subsection{Alternative (a1): {${\mathbf{n_c < n_b + 2< n_b+3 < n_s}}$}}

At this stage, the further breaking of the flavour symmetry depends
on the details about the assumed power counting for the quark masses.
Let us first discuss the scenario (a1):

\subsubsection{Order $\Lambda''/\Lambda$}

In the case $n_c < n_b+2$, it is convenient
to classify the residual spurion fields of $G_F''$ 
according to (\ref{lamp:2}),
\begin{align}
Y_U^{(2)} & \sim (2,2,1)_{0,-1,0} 
\ \propto \ 
\left( \begin{array}{cc} 
\lambda^{1+n_u} & \lambda^{1+n_c}  \\
\lambda^{1+n_u} & \lambda^{n_c}  
\end{array} \right) \Lambda 
\,, \cr 
Y_D^{(2)} & \sim (2,1,2)_{0,0,-1}
\ \propto \
\left( \begin{array}{cc} 
\lambda^{n_d} & \lambda^{1+n_s} \\
\lambda^{1+n_d} & \lambda^{n_s} 
\end{array} \right) \Lambda 
\,, \cr 
\chi_s & \sim (2,1,1)_{-1,0,0} 
\propto \left( \begin{array}{c} 
\lambda^{3+n_b} \\
\lambda^{2+n_b} 
\end{array} \right) \Lambda \,,
\end{align}
such that the next spurion getting a VEV is
\begin{align}
 \langle Y_U^{(2)} \rangle = 
\left( \begin{array}{cc} 
0 & 0  \\
0 & \tilde y_c  
\end{array} \right) \Lambda 
\equiv
\left( \begin{array}{cc} 
0 & 0  \\
0 & x_c  
\end{array} \right) \Lambda''
\end{align}
with $x_c = {\cal O}(1)$, implying $\tilde y_c\sim \Lambda''/\Lambda \sim m_c/m_t$.
The VEV of $Y_U^{(2)}$ thus further breaks the flavour symmetry as
\begin{align}
G_F'' \to G_F^{(3a1)}
& = SU(2)_{D_R} \times U(1)_{C} \times  U(1)_{\rm III} 
 \times U(1)_{Q_L^{(1)}} \times U(1)_{U_R^{(1)}} 
\label{lam3a:1}
\\
 & \sim SU(2)_{D_R} \times U(1)_{C} \times  U(1)_{Q_L^{(1)}} \times U(1)_{U_R^{(1)}} \times U(1)_{D_R^{(2)}}
\label{lam3a:2}
 \,,
\end{align}
where $U(1)_C$ refers to the second-generation quarks in $Q_L$ and $U_R$.
This implies 5 additional Goldstone bosons ($\Pi_{L,U_R}''{}^{a=1..3}$, $\Pi''_L{}^3=-\Pi''_{U_R}{}^3$), appearing via
\begin{align}
Y_U^{(2)}  & = 
{\cal U}(\Pi''_L)
\left( \begin{array}{cc} 
Y_U^{(1)}  & 0  \\
0 & x_c \, \Lambda'' 
\end{array} \right) {\cal U}^\dagger(\Pi_{U_R}'')  \,, \\
Y_D^{(2)} & = {\cal U}(\Pi''_L)
\left( \begin{array}{c}
        \xi_d^\dagger \\ \xi_s^\dagger
       \end{array} \right)
 \,,
\quad
\chi_s = {\cal U}(\Pi''_L)
\left( \begin{array}{c} \chi_{13} \\ \chi_{23} \end{array} \right) \,.
\label{chis}
\end{align}

\subsubsection{Order $\Lambda^{(3)}/\Lambda$}

The residual spurions of $G_F^{(3a1)}$ now scale/transform as
\begin{align}
 Y_U^{(1)} & \sim (1)_{0,1,-1,0} \ \propto \ \lambda^{n_u} \,,
 \cr 
 \xi_d^\dagger & \sim (2)_{0,1,0,-1} \ \propto \left( \begin{array}{cc} 
\lambda^{n_d} & \lambda^{1+n_s} \end{array} \right) \,, 
\qquad  \chi_{13}  \sim (1)_{0,1,0,0} \ \propto \ \lambda^{3+n_b} \,, 
\cr 
 \xi_s^\dagger & \sim (2)_{1,0,0,-1} \ \propto \left( \begin{array}{cc} 
\lambda^{1+n_d} & \lambda^{n_s} \end{array} \right) \,, 
\qquad 
 \chi_{23}  \sim (1)_{1,0,0,0} \ \propto \ \lambda^{2+n_b} \,,
\end{align}
where the subscripts refer to the $U(1)$ charges in (\ref{lam3a:2}).
In this case, assuming $n_s > n_b +2$,
the next spurion to receive a VEV is $\chi_{23}$, 
which breaks the $U(1)_C$ symmetry,
\begin{align}
 G_F^{(3a1)}
\to G_F^{(4a1)} & 
=  SU(2)_{D_R} \times U(1)_{Q_L^{(1)}} \times U(1)_{U_R^{(1)}} \times U(1)_{D_R^{(2)}} \,.
\end{align}
The associated Goldstone boson $\phi'''$ appears as a simple phase, 
\begin{align}
 \chi_{23} = x_{sb} \, e^{i \phi'''/\Lambda'''} \, \Lambda'''
\qquad \mbox{and} \qquad \xi_s^\dagger \to \xi_s^\dagger \, e^{i \phi'''/\Lambda'''}
\end{align}
with $\Lambda'''/\Lambda \sim y_b \lambda^2$.

\subsubsection{Order $\Lambda^{(4)}/\Lambda$}

The residual spurions for $G_F^{(4a1)}$ read 
\begin{align}
 Y_U^{(1)} & \sim (1)_{1,-1,0} \ \propto \ \lambda^{n_u} \,, \cr 
 \xi_d^\dagger & \sim (2)_{1,0,-1} \ \propto \left( \begin{array}{cc} 
\lambda^{n_d} & \lambda^{1+n_s} \end{array} \right) \,, \cr   
 \xi_s^\dagger & \sim (2)_{0,0,-1} \ \propto \left( \begin{array}{cc} 
\lambda^{1+n_d} & \lambda^{n_s} \end{array} \right)  \,,
\cr 
 \chi_{13}  & \sim (1)_{1,0,0} \ \ \, \propto \ \lambda^{3+n_b} \,. 
\end{align}
For $n_s > n_b +3$, the next spurion to get a VEV is 
$\chi_{13}$ which breaks another $U(1)$ symmetry,
\begin{align}
 G_F^{(4a1)} \to G_F^{(5a1)} = SU(2)_{D_R} \times U(1)_{U_R^{(1)}} \times U(1)_{D_R^{(2)}}  \,.
\end{align}
The corresponding Goldstone mode $\phi^{(iv)}$ appears again as a phase factor used
to redefine the spurions $Y_U^{(1)}$  and $\xi_d$ according to their $U(1)$ charge.
It should be noted that after the $U(1)_{Q_L^{(1)}}$ is broken,
the residual spurions $\xi_d$ and $\xi_s$ have the same quantum numbers
with respect to $G_F^{(5a1)}$, and therefore the relative phase of their VEVs
will provide the source for spontaneous CP-violation.

\subsubsection{Order $\Lambda^{(5)}/\Lambda$}

Taking $\langle \xi_s \rangle \neq 0$, we next break
\begin{align}
  G_F^{(5a1)} \to G_F^{(6a1)} = U(1)_{U_R^{(1)}} \times U(1)_{D_R^{(1)}}
\end{align}
and the remaining spurion fields are given by
\begin{align}
   Y_U^{(1)} & \sim (1)_{-1,0} \propto \lambda^{n_u} \,, \quad
   Y_D^{(1)}  \sim (1)_{0,-1} \propto \lambda^{n_d} \,, \quad 
   \chi_{12}  \sim (1)_{0,0} \propto \lambda^{1+n_s} \,.
\end{align}
Here we have decomposed the $SU(2)_{D_R}$ doublet $\xi_d$
into two complex singlets $Y_D^{(1)}$ and $\chi_{12}$,
\begin{align}
 \xi_s^\dagger &= (0,\ x_{ss} \, \Lambda^{(5a1)}) \ {\cal U}^\dagger(\Pi^{(v)}_R) \,, 
\qquad 
 \xi_d^\dagger = (Y_D^{(1)} ,\ \chi_{12} ) \ {\cal U}^\dagger(\Pi^{(v)}_R) \,.
\end{align}

\subsection{Alternative (a2): {${\mathbf{n_c < n_b + 2< n_s < n_b+3}}$}}

\subsubsection{Order $\Lambda''/\Lambda$ and Order $\Lambda^{(3)}/\Lambda$}

These steps are the same as for the alternative (a1) above.

\subsubsection{Order $\Lambda^{(4)}/\Lambda$}

In this case, i.e.\ for $n_s < n_b + 3$,
the next spurion to receive a VEV is $\langle \xi_s
\rangle \sim y_s \Lambda$, 
which breaks
\begin{align}
 G_F^{(4a1)} \to G_F^{(5a2)} = U(1)_{Q_L^{(1)}} \times U(1)_{U_R^{(1)}} \times U(1)_{D_R^{(1)}}
\end{align}
leaving us with 4 singlet spurion fields
\begin{align}
 Y_U^{(1)} &\sim (1)_{1,-1,0}  \propto \lambda^{n_u} \,, \cr 
 Y_D^{(1)} &\sim (1)_{1,0,-1}  \propto \lambda^{n_d} \,, \cr 
 \chi_{12} &\sim (1)_{1,0,0}  \propto \lambda^{n_s+1} \,, \cr
 \chi_{13} &\sim (1)_{1,0,0}  \propto \lambda^{n_b+3} \,. 
\end{align}

\subsubsection{Order $\Lambda^{(5)}/\Lambda$}

Taking now $\langle \chi_{13} \rangle \neq 0$, we break
\begin{align}
  G_F^{(5a2)} \to G_F^{(6a2)} = U(1)_{U_R^{(1)}} \times U(1)_{D_R^{(1)}}
\end{align}
with the remaining spurion fields as for case (a1).

\subsection{Alternative (b): {${\mathbf{n_b + 2< n_c < n_s < n_c + 1}}$}}

The case $n_c > n_b + 2$ may be considered as somewhat less likely, because
in order to have $y_c/y_b \lesssim \lambda^2$ at some high scale we would have
to require sizeable renormalization effects in order to recover $m_c/m_b \sim 0.3$
at low scales.

\subsubsection{Order $\Lambda''/\Lambda$}

In that case, the next spurion to get a VEV would be 
\begin{align} 
 \langle \chi_s \rangle &= 
 \left( \begin{array}{c} 0 \\ y_{23} \end{array} \right) \Lambda = 
 \left( \begin{array}{c} 0 \\ x_{23} \end{array} \right) \Lambda'' 
\end{align}
with $x_{23} = {\cal O}(1)$ and thus $\Lambda''/\Lambda = \lambda^2 m_b/m_t$.
This leads to the breaking
\begin{align} 
G_F'' &\to G_F^{(3b)}
 = SU(2)_{U_R} \times SU(2)_{D_R} 
 \times U(1)_{Q_L^{(1)}} \times U(1)_{U_R^{(2)}}\times U(1)_{D_R^{(2)}} \,.
\end{align}
Introducing three new Goldstone bosons ($(\Pi_L'')^{a=1,2,3}$), we parametrize
\begin{align} 
& 
Y_U^{(2)}  = {\cal U}(\Pi_L'') \, 
 \left( \begin{array}{c} \xi_u^\dagger \\ \xi_c^\dagger \end{array} \right) \,,
\quad
Y_D^{(2)}  = {\cal U}(\Pi_L'') \, 
 \left( \begin{array}{c} \xi_d^\dagger \\ \xi_s^\dagger \end{array} \right) \,,
\quad
\chi_s  = {\cal U}(\Pi_L'') \, \langle \chi_s \rangle \,.
\end{align}

%%%%%%%%%%%%%%%%%%%%%%%%%%%%%%%%%%%%%%%%%%%%%%%%%%%%%%%%%%%%%%%%%%%%%%%%%%%%%%%

\subsubsection{Order $\Lambda^{(3)}/\Lambda$}

The residual spurions of $G_F^{(3b)}$ scale/transform as
\begin{align}
  \xi_u^\dagger & \sim (2,1)_{1,-1,0} 
      \ \propto \left( \begin{array}{cc} \lambda^{n_u}& \lambda^{1+n_c} \end{array} \right) \,, \qquad  
\xi_d^\dagger  \sim (1,2)_{1,0,-1} 
      \ \propto\left( \begin{array}{cc} \lambda^{n_d} & \lambda^{1+n_s} \end{array} \right) \,,
\cr 
   \xi_c^\dagger & \sim (2,1)_{0,-1,0} 
      \ \propto \left( \begin{array}{cc} \lambda^{1+n_u} & \lambda^{n_c} \end{array} \right)\,,  \qquad 
  \xi_s^\dagger  \sim (1,2)_{0,0,-1}  
      \propto\left( \begin{array}{cc} \lambda^{1+n_d} & \lambda^{n_s}\end{array} \right) \,.
\end{align}
In this case, the next spurion to receive a VEV is $\xi_c^\dagger$,
which breaks
\begin{align}
 G_F^{(3b)} \to G_F^{(4b)} 
 & = SU(2)_{D_R} 
 \times U(1)_{Q_L^{(1)}} \times U(1)_{U_R^{(1)}}\times U(1)_{D_R^{(2)}} \,,
\end{align}
introducing three new Goldstone bosons at the scale $\Lambda^{(3)} \sim y_c \Lambda$.

\subsubsection{Order $\Lambda^{(4)}/\Lambda$}

Decomposing the doublet $\xi_u^\dagger$ into two singlets $Y_U^{(1)}$ and $\varphi_{12}$,
the remaining spurions of $G_F^{(4b)}$ are
\begin{align}
 Y_U^{(1)} &\sim (1)_{1,-1,0} \ \propto \ \lambda^{n_u} \,,\qquad
\mbox{and} \qquad  \varphi_{12} \sim (1)_{1,0,0} \ \propto \ \lambda^{1+n_c} \,, \cr 
  \xi_d^\dagger & \sim (2)_{1,0,-1} 
      \ \propto \ \left( \begin{array}{cc} \lambda^{n_d} & \lambda^{1+n_s} \end{array} \right) \,, \cr  
  \xi_s^\dagger  & \sim (2)_{0,0,-1}  
      \ \propto \ \left( \begin{array}{cc} \lambda^{1+n_d} & \lambda^{n_s}\end{array} \right) \,.
\end{align}
Notice that the flavour group and the representations of the spurion fields
are the same as for $G_F^{(4a1,4a2)}$, only that the role of $\chi_{13}$ is now
played by $\varphi_{12}$.

As in the case of scenario (a2), we assume that the next spurion to get a 
VEV is $\xi_s$, breaking the flavour symmetry at $\Lambda^{(4b)} \sim y_s \Lambda$,
\begin{align}
 G_F^{(4b)}
\to G_F^{(5b)} & 
=  U(1)_{Q_L^{(1)}} \times U(1)_{U_R^{(1)}} \times U(1)_{D_R^{(1)}} \,.
\end{align}
The remaining steps in the flavour symmetry breaking follow scenario (a2),
except for $\Lambda^{(5b)} \sim \langle \varphi_{12} \rangle \sim \lambda y_c \Lambda$.

\subsection{Order $\Lambda^{(6)}/\Lambda$ and Order $\Lambda^{(7)}/\Lambda$}

Since in all scenarios the residual spurion field $\chi_{12}$
is uncharged under $G_F^{(6)}$, its VEV will in general be a complex
number whose phase cannot be rotated away by flavour transformations.
The CP symmetry in the Yukawa sector will thus be 
broken spontaneously by $\langle \chi_{12} \rangle/\Lambda \sim \lambda y_s \sim \Lambda^{(5)}/\Lambda$, if the potential singles out a non-vanishing imaginary part.

Finally, the VEVs for $Y_U^{(1)}$ and $Y_D^{(1)}$ break the remaining
flavour symmetry
\begin{align}
 G_F^{(6a1)} & \to \mbox{nothing}
\end{align}
and give masses to the up- and down-quark, where the order
of symmetry breaking is not really important.

\section{Various $U(1)$ charges}
\label{app:u1}
For convenience, we collect in Table~\ref{tab:u1} the various
$U(1)$ charges appearing in the construction of the flavour symmetry
breaking. Notice that some $U(1)$ charges are linear dependent,
\begin{align}
3B &= T + Q_L^{(2)} + U_R^{(2)} + D_R  \,,
\\
& = {\rm III} + Q_L^{(2)} + U_R^{(2)} + D_R^{(2)} \,,
\\
& = {\rm III} + C + Q_L^{(1)} + U_R^{(1)} + D_R^{(2)} \,.
\end{align}

\begin{table}
 \small
\caption{\label{tab:u1} Various $U(1)$ charges appearing in the discussion of the
sequential flavour symmetry breaking.}
\begin{center}
\begin{tabular}{|c||c c c||c c| c c c || c | c c c|}
\hline
          & $3B$ & $U_R$ & $D_R$ & T & III & $Q_L^{(2)}$ & $U_R^{(2)}$ & $D_R^{(2)}$  &
            $C$ &  $Q_L^{(1)}$ & $U_R^{(1)}$ & $D_R^{(1)}$
\\
\hline \hline 
$(u,d)_L$ & 1   &  0    &  0    &  0 & 0 & 1  & 0        & 0            &  
            0   &  1    &  0    &  0
\\
$(c,s)_L$ & 1   &  0    &  0    &  0 & 0 & 1  & 0        & 0            & 
            1   &  0    &  0    &  0
\\
$(t,b)_L$ & 1   &  0    &  0    &  1 & 1 & 0  & 0        & 0            & 
            0   &  0    &  0    &  0
\\
\hline
$u_R$     & 1   &  1    &  0    &  0 & 0 & 0  & 1        & 0            &  
            0   &  0    &  1    &  0
\\
$c_R$     & 1   &  1    &  0    &  0 & 0 & 0  & 1        & 0            & 
            1   &  0    &  0    &  0
\\
$t_R$     & 1   &  1    &  0    &  1 & 1 & 0  & 0        & 0            &  
            0   &  0    &  0    &  0
\\
\hline
$d_R$     & 1   &  0    &  1    &  0 & 0 & 0  & 0        & 1            &  
            0   &  0    &  0    &  1
\\
$s_R$     & 1   &  0    &  1    &  0 & 0 & 0  & 0        & 1            & 
            0   &  0    &  0    &  0
\\
$b_R$     & 1   &  0    &  1    &  0 & 1 & 0  & 0        & 0            &  
            0   &  0    &  0    &  0
\\
\hline
\end{tabular}
\end{center}
\end{table}

\section{Expressing the $G_F$-invariants through $G_F^\prime$-invariants}

\label{app:jinv}

The explicit relations between the 10 invariants $i_\alpha^{(m)}$ of the full flavour
group $G_F$ and the 9 invariants $i_\alpha^{(m)}$ of the residual flavour group
$G_F'$ read
\begin{align}
& i_1^{(2)}   = j_1^{(2)} + y_t^2 \, \Lambda^2 \,,
\qquad
 i_2^{(2)} = j_2^{(2)} + j_3^{(2)} \,,
\end{align}
for the dimension-2 invariants, and
\begin{align}
& i_1^{(4)} = j_1^{(4)} - 2 \, y_t^2 \Lambda^2  j_1^{(2)} \,, 
\cr
& i_2^{(4)} = j_2^{(4)} - j_1^{(2)} j_3^{(2)} - y_t^2  \Lambda^2 j_2^{(2)} \,,
\cr
& i_3^{(4)} = j_3^{(4)} + 2 j_4^{(4)}  \,,
\end{align}
for the dimension-4 terms, together with
\begin{align}
 i_1^{(6)} & = -\frac32 \, y_t^2\Lambda^2 j_1^{(4)} \,,
\cr 
 i_2^{(6)} &= - y_t^2\Lambda^2 j_2^{(4)} - \frac12\, j_3^{(2)} j_1^{(4)}
\cr 
 i_3^{(6)} &= j_1^{(6)}- \frac12 \, y_t^2\Lambda^2 j_3^{(4)} 
 - j_1^{(2)} \left( j_4^{(4)} +j_2^{(2)}J_3^{(2)} \right) - j_3^{(2)} j_2^{(4)} \,,
\cr 
 i_4^{(6)} &= 3 j_2^{(6)} 
 - 3 j_2^{(2)} \left( j_4^{(4)} + j_2^{(2)}j_3^{(2)} \right)
 - \frac32 j_3^{(2)} j_3^{(4)} \,,
\end{align}
and
\begin{align}
 i_1^{(8)} & = \left( j_1^{(2)} 
  - 2 y_t^2 \Lambda^2 \right) j_1^{(6)} 
  + \left( y_t^4\Lambda^4 + \frac12 j_1^{(4)} \right)
     \left(j_4^{(4)}+ j_2^{(2)}J_3^{(2)} \right)
\cr & \quad 
+ j_1^{(4)} j_3^{(4)}
- j_2^{(4)} \left( j_2^{(4)} + j_1^{(2)} j_2^{(2)} \right)
+\frac12 (j_1^{(2)})^2 j_3^{(4)}
+\frac12 (j_2^{(2)})^2 j_1^{(4)}
\,.
\end{align}

\end{appendix}

\end{document}